%% file: main.tex
\documentclass[conference,10pt]{IEEEtran}

\ifCLASSINFOpdf
\else
\fi

\usepackage{amsmath}
\usepackage{url}
\usepackage{pgfplots}
\usepgfplotslibrary{groupplots}
\pgfplotsset{compat=1.18}
\usepackage{pifont}
\newcommand{\cmark}{\ding{51}}
\newcommand{\xmark}{\ding{55}}

\usepackage{wasysym}
\newcommand{\full}{\CIRCLE}      
\newcommand{\half}{\LEFTcircle}  

\usepackage{multirow}
\usepackage{comment}
\usepackage{array}
\usepackage{cleveref}
\usepackage{booktabs} 
\usepackage{ragged2e} 
\newcolumntype{P}[1]{>{\RaggedRight\arraybackslash}p{#1}} 
\usepackage{tikz}
\usetikzlibrary{shapes.geometric, arrows.meta, positioning, fit, backgrounds}
\usepackage{multirow}
\usepackage[table,xcdraw,dvipsnames]{xcolor}

\definecolor{tbone}{HTML}{4D9DE0}
\definecolor{tbtwo}{HTML}{E15554}
\definecolor{lwone}{HTML}{F6AA1C}
\definecolor{lwtwo}{HTML}{3BB273}
\definecolor{lwthree}{HTML}{7768AE}

\usepackage{makecell}
\usepackage{siunitx}

\newcommand{\drawkey}[3]{
    \draw[line width=0.07cm, draw=#2] (#1) circle [radius=0.15cm];
    \draw[line width=0.07cm, draw=#2] (#1 -0.15) -- ++(0,-0.4);
    \draw[line width=0.07cm, draw=#2] (#1 -0.35) -- ++(-0.2,0);
    \draw[line width=0.07cm, draw=#2] (#1 -0.51) -- ++(-0.2,0);
    \node[below, text=#2] at (#1 -0.6) {#3}
}

\begin{document}

    \title{Efficient and Quantum-safe Internet Key Exchange Protocols for Satellite Communications}

    \author{\IEEEauthorblockN{Davide De Zuane}
            \IEEEauthorblockA{\textit{IMT School for Advanced Studies}\\ Lucca, Italy}
            \and 
            \IEEEauthorblockN{Marco Baldi, Paolo Santini}
            \IEEEauthorblockA{\textit{Università Politecnica delle Marche}\\ Ancona, Italy}
            \and 
            \IEEEauthorblockN{Grégoire Anchelergues, Daniele Romano\\ Alessandro Cammarano, Juan José Grosso}
            \IEEEauthorblockA{\textit{OSMIUM}, Turin, Italy}

        \thanks{This work was partially supported by the European Space Agency (ESA ESTEC) under the call AO/1-11711/23/NL/FGL (RE-ISSUE) - Lightweight post-quantum key exchange protocol for IP data transfers over satellite, project “SATEllite Lightweight Internet Key Exchange” (SATELIKE) and by the Italian Ministry of University and Research (MUR) under the Italian Fund for Applied Science (FISA 2022), Call for tender No. 1405 published on 13-09-2022 - project title “Quantum-safe cryptographic tools for the protection of national data and information technology assets” (QSAFEIT) - No. FISA 2022-00618 (CUP I33C24000520001), Grant Assignment Decree no. 15461 adopted on 02.08.2024.}
    }

    \IEEEoverridecommandlockouts
    \maketitle

    \begin{abstract}
        This paper studies cryptographic key exchange in satellite communications, which requires specific solutions because the satellite context presents unique challenges, particularly concerning onboard resource constraints and long transmission latency. We address these challenges by considering the Internet Key Exchange (IKE) protocol, which is widely used in terrestrial networks, and studying its applicability in the satellite context. This requires addressing two main issues: i) its efficiency in terms of the resources and bandwidth required to adapt to satellite terminals, and ii) its resistance even to attackers equipped with a quantum computer, in order to resist obsolescence and defend against harvest-now-decrypt-later attacks. We study these aspects from both a design and experimental point of view, defining and assessing some protocol variants characterized by low complexity and quantum resistance. To address the need to manage the transition from classic cryptographic primitives to post-quantum ones, we also consider the possibility of using hybrid cryptographic solutions that combine them both.
    \end{abstract}


    \IEEEpeerreviewmaketitle

    \section{Introduction}
    
    The protection of both inter-satellite and satellite-to-ground communications has become a critical concern, requiring the use of encryption to ensure confidentiality of the transmitted data. Furthermore, the growing complexity of satellite networks requires automated mechanisms for the distribution of encryption keys based on asymmetric cryptography -- a paradigm that, in its current form, is vulnerable to quantum computer-based attacks. The sensitive nature of the data transmitted across these links, combined with the long operational lifetimes of satellite systems, thus requires the adoption of post-quantum cryptographic techniques capable of withstanding attacks from both classical and quantum adversaries. In this paper, we propose and assess a solution to this problem considering the Internet Key Exchange (IKE) protocol and some new variants of it specifically designed for the satellite environment. In fact, adapting the IKE protocol for satellite communications involves addressing two main challenges:
        
    \begin{itemize}
        \item Optimizing the protocol in terms of complexity and latency.
        \item Integrating post-quantum asymmetric cryptographic primitives to withstand attacks based on quantum computers, including harvest-now-decrypt-later attacks.
    \end{itemize}

    We define two baseline variants and three lightweight variants of the IKEv2 protocol for use in satellite networks, aimed at progressively reducing computational and communication overhead, and at introducing post-quantum asymmetric cryptographic primitives, possibly in a hybrid combination with classical ones. The performance of the proposed variants is evaluated through simulations, taking into account some possible configurations of the satellite network, which allow for an assessment of the cost-benefit ratio of each variant.

    \section{Background and related works}

    IPsec (Internet Protocol Security) is a set of standard protocols that provide mechanisms for authentication, encryption and integrity of data transmitted between two or multiple devices, thus protecting IP communications from eavesdropping and tampering. 
    The main components of IPsec are the ESP (Encapsulating Security Payload) and AH (Authentication Header) protocols for data authentication and confidentiality, SAs (Security Associations) for the definition of cryptographic functions and sets of parameters and the IKE (Internet Key Exchange) protocol for negotiation of SA parameters, authentication and dynamic key agreement.

    \begin{figure}[t!]
        \centering
        \input{figures/ike_exchange}
        \caption{Main IKEv2 protocol exchanges. Messages in 
            \textcolor{ForestGreen}{green} are sent by the initiator, messages in \textcolor{Blue}{blue} by the responder. The derived session key \texttt{SK\_d} is shown in \textcolor{orange}{orange}.
        }
        \label{fig:ike_exchange}
    \end{figure}
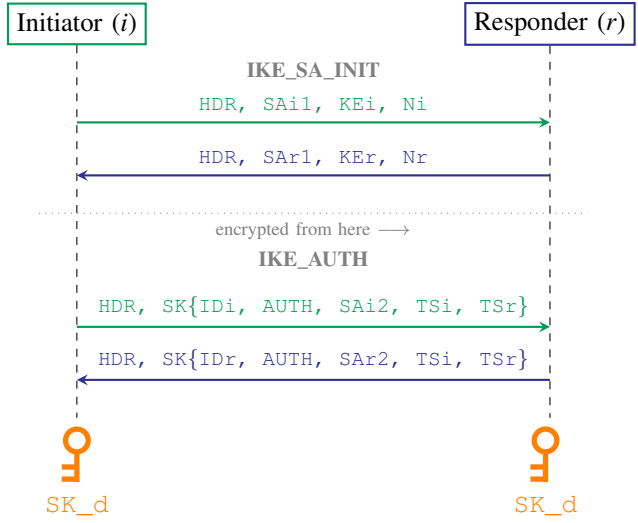

    The IKE protocol is responsible for the initial establishment of a secure connection between two endpoints, performing the functions of authentication and the exchange of the initial cryptographic material needed to protect the rest of the communication. It therefore covers the phase in which asymmetric cryptography must be used to perform authentication and key exchange between the two endpoints over an insecure communication channel such as the Internet.
    This protocol is of particular interest in satellite networks as well, where historically each connection was secured through an initial pre-sharing of symmetric keys, a practice that however is no longer feasible in the context of modern satellite networks, including satellite mega-constellations, which are increasingly resembling a space-based analog of the Internet. The most recent and secure version of IKE is IKEv2, which adopts an initiator-responder model where communication is established through two phases -- \texttt{IKE\_SA\_INIT} and \texttt{IKE\_AUTH} -- as shown in \Cref{fig:ike_exchange}.
    The meanings of the abbreviations reported in the figure and the content of the corresponding messages are explained in \Cref{tab:ike_payloads}.

    \input{tables/ike_payloads}

    Integrating post-quantum primitives into IKEv2 is not straightforward. The protocol was designed around compact classic cryptographic tools, and the structural limitations that emerge when attempting this transition -- including payload size growth, fragmentation handling, and the lack of native support for large key material -- have been analyzed in our prior work~\cite{DeZuane2026}. When the target deployment is a satellite network, additional constraints come into play: link latency makes every extra round trip costly, communication windows may be limited, and the long operational lifetime of satellites introduces long-term confidentiality requirements that classical algorithms alone cannot meet. Furthermore, satellite hardware is typically resource-constrained -- limited in processing power, memory, and energy budget -- as payloads are optimized for size and weight rather than computational capacity. This makes the adoption of post-quantum primitives, which are generally more demanding than their classical counterparts in terms of both computation and memory footprint, a non-trivial challenge even beyond the protocol-level considerations. Furthermore, although standardized post-quantum cryptographic primitives now exist, their integration into standard versions of the IKE protocol is still largely lacking. RFC~8784~\cite{rfc8784} provides post-quantum pre-shared key support as an interim mitigation against harvest-now-decrypt-later attacks, while RFC~9370~\cite{rfc9370} generalizes the key exchange framework to support hybrid cryptography, though without addressing the challenges posed by post-quantum key sizes. The only document directly targeting a post-quantum primitive in IKEv2 is an IETF draft~\cite{ietf-ipsecme-ikev2-mlkem-04} proposing the use of ML-KEM in the \texttt{INIT} exchange, leaving authentication and the broader transition largely unaddressed. There are also few scientific studies addressing the post-quantum transition of the IKE protocol. In addition to our previous work \cite{DeZuane2026} on the variant known as Minimal IKE \cite{rfc7815}, the work \cite{Mutlugun2024} studies the introduction of post-quantum cryptographic primitives into IKEv2, including a simulation of its performance over satellite networks. However, only the fully-fledged standard version of the IKEv2 protocol is considered, without exploring how to optimize it for a satellite environment or possible variants using hybrid encryption.

    \section{Making IKEv2 lightweight and quantum-safe}

    The use of IPsec introduces overhead both in terms of the bandwidth required for communication and the computational capacity required at the terminals.
    First, we examine some techniques that we can use to minimize overheads in scenarios with limited bandwidth. Then, we address the problem of computationally constrained endpoints. A first step in this direction is to consider simplified versions of the IKEv2 protocol, such as Minimal IKE \cite{rfc7815}. In bandwidth-constrained satellite environments, reducing protocol overhead is essential to reduce latency and achieve a lightweight IKEv2/IPsec deployment. The encapsulation introduced by ESP or AH -- particularly in tunnel mode -- adds additional IP headers and cryptographic metadata (e.g., IVs, padding, integrity tags), which decrease \textit{goodput}. This problem can be mitigated by resorting to header and payload compression techniques, like the IP Payload Compression Protocol (IPComp)~\cite{rfc2393}, which reduces payload size and is widely supported, though its effectiveness varies by traffic type, and the Robust Header Compression (ROHC)~\cite{rfc5795} protocol, targeting protocol headers.

    \subsection{Integration of post-quantum cryptography}
    \label{sec:combiner}

    There is an ongoing debate over whether to entirely replace quantum-vulnerable cryptographic primitives with post-quantum alternatives, or to combine them in a hybrid approach to avoid relying entirely on relatively new cryptographic primitives. In the second case, key negotiation using two different primitives could be handled natively through a double exchange, but this clearly has a significant impact on the protocol's performance. A similar argument applies to the authentication phase, where post-quantum digital signatures introduce analogous overhead. In both cases, the same combiner-like philosophy can be applied, though the specific constraints differ between the two exchanges. This approach has been explored in the context of other protocols as well: \cite{stebila-prototyping-ssh-tls} investigates hybrid key exchange and authentication in TLS and SSH, providing evidence that such combinations are practically viable.

    \subsubsection{Key Exchange}

    In its standard form, RFC~9370 \cite{rfc9370} achieves hybrid key exchange by building on the \textit{IKE Intermediate Exchange}, requiring one additional round trip per extra key exchange. In satellite communications, where link latency is already significant, this overhead carries a tangible operational cost. In deployments where backward compatibility is not a requirement, a \textit{combiner-like} approach based on payload concatenation can be employed instead. The classical and post-quantum key exchanges are carried out within the existing \texttt{INIT} exchange, eliminating the extra round trips. However, the \texttt{INIT} exchange carries an additional constraint: its messages must be MTU-safe, as the \textit{IKE Fragmentation} extension is not yet active at this stage of the handshake. This imposes a practical bound on which post-quantum algorithms can be accommodated within a single \texttt{INIT} message, and algorithm selection must take this constraint into account.

    \subsubsection{Authentication}

    An analogous challenge arises in the \texttt{AUTH} phase. The standard solution again relies on the \textit{IKE Intermediate Exchange} to carry this additional material at the cost of extra round trips. In constrained deployments, the combiner-like approach can be applied here as well: classical and post-quantum authentication material are combined within the existing \texttt{AUTH} exchange, avoiding the extra-RT while retaining the security benefits of hybrid authentication. Importantly, by the time the \texttt{AUTH} exchange takes place, the \textit{IKE Fragmentation} extension is already active, without risking IP-level fragmentation. This makes the \texttt{AUTH} phase more accommodating than \texttt{INIT} for the inclusion of post-quantum material, and relaxes the constraint on algorithm selection accordingly. A further degree of freedom is offered by RFC~7427~\cite{rfc7427}, which generalizes the signature authentication mechanism in IKEv2 to support arbitrary digital signature algorithms. By leveraging this extension, it becomes possible to avoid transmitting large post-quantum public keys within the exchange altogether: the public key can instead be pre-shared via an out-of-band mechanism, and RFC~7427 allows the peer to reference and use it directly during authentication.

    \section{IKEv2 variants for satellite communications}
    
    The previous section examined the extensions and modifications available to reduce the overhead of IKEv2 and make it quantum-safe. Translating these considerations into a concrete implementation, however, is a non-trivial task. IKEv2 has accumulated a large number of extensions over the years, and achieving a coherent, optimized implementation that selectively incorporates only the relevant ones requires careful engineering effort. We consider strongSwan\footnote{\url{https://strongswan.org/}}, one of the most widely adopted and actively maintained IKEv2 implementations, as our starting point. Building on it, we realize the optimizations discussed in the previous section through two stages of refinement. The first defines a \textit{technical baseline}, representing the maximum optimization achievable within the bounds of currently standardized and supported features. The second introduces \textit{lightweight variants} that go beyond this baseline by incorporating features not yet available in strongSwan -- as well as the protocol-level modifications considered in this work -- to address the specific constraints of satellite communications. Although these constraints primarily concern latency, the entire protocol architecture has been revised to retain only those functions that are required in the satellite context under consideration. The protocol variants we are considering are described next, and a summary of their features is provided in Table \ref{tab:variants_summary}.
  
    \input{tables/ike_variants}

    \subsection{TB1: Plain IKEv2}
    
    This variant represents the maximum optimization achievable within the current capabilities of strongSwan, without introducing any modifications to the protocol or its implementation. 
    The adopted cipher suite includes an additional key exchange algorithm that requires an additional round trip. Authentication relies on raw public keys, which must be transmitted over the network during the \texttt{AUTH} exchange. This already introduces non-negligible bandwidth overhead, which would be further exacerbated by the use of public key certificates, as these carry additional metadata and chain information on top of the public key material itself. 
    Crucially, this variant offers no post-quantum protection for the authentication phase.

    \subsection{TB2: Minimal IKEv2 with Pre-Shared Keys }
    \label{sec:tb2}

    By combining ML-KEM-768 for key exchange with pre-shared keys (PSKs) for authentication, according to RFC~8784~\cite{rfc8784}, this variant achieves full post-quantum security for both phases within a compact two-message handshake, without incurring the additional round trip required by RFC~9370. However, this variant comes with some limitations. PSK-based authentication requires keys to be distributed out-of-band prior to protocol execution, which does not scale well in large or dynamic networks. The absence of hybrid cryptography further means that security relies entirely on the post-quantum algorithm, with no classical fallback in case of unforeseen cryptanalytic advances. Finally, the choice of ML-KEM-768 over higher security levels is not arbitrary: the MTU constraint on the \texttt{INIT} exchange directly caps the size of the key exchange payload, and therefore the achievable post-quantum security level. This is an intrinsic limitation of any configuration that avoids the \texttt{INTERMEDIATE} exchange.

    \subsection{LW1: Full-Capability Mode}

    LW1 represents a first lightweight protocol variant we propose, extending IKEv2 with hybrid key exchange and post-quantum authentication. The former is implemented according to RFC~9370. In the authentication phase, instead, some design choice is required. In fact, unlike key exchange, for which hybrid approaches are well established and standardized through RFC~9370, the question of how to hybridize authentication in IKEv2 remains open. In the absence of a standardized hybrid authentication mechanism, we follow the simplest approach, replacing the classical signature scheme with a post-quantum one directly. This results in a \textit{partial hybridization}: the key exchange benefits from the security guarantees of both classical and post-quantum cryptography, while authentication relies solely on a post-quantum primitive. There are two possible approaches to distributing the corresponding public keys: raw public keys or X.509 certificates (in both cases, public key authentication is ensured through out-of-band mechanisms). While raw public keys are more compact, their integration requires an additional protocol extension. X.509 certificates, instead, act as a self-contained object for the public key material and allow the protocol to remain fully compliant with the existing IKEv2 specification, without any additional trick. For this reason, we adopt X.509 certificates for the distribution of public keys in this variant. In either case, the associated bandwidth overhead is entirely absorbed by pre-sharing the material via out-of-band mechanisms prior to protocol execution. As a result, no key or certificate is transmitted in-band during the exchange, and the size of \texttt{AUTH} payloads is kept minimal.

    \subsection{LW2: Minimal Overhead Mode}

    LW2 pushes the minimization further by adopting the combiner-based hybridization strategy described in Section~\ref{sec:combiner}. Classical and post-quantum payloads are concatenated and embedded directly within a unique message. This minimizes the overhead of the handshake procedure  while retaining full hybrid cryptography in both phases. In fact, differently from LW1, where authentication relies solely on a post-quantum signature scheme, LW2 adopts a fully hybrid authentication approach: both a classical and a post-quantum signature are computed and combined within the \texttt{AUTH} exchange. The downside is that this approach compromises backward compatibility with existing implementations of the IKEv2 standard protocol. This, however, is a deliberate trade-off: the goal is to minimize the number of exchanges and the associated latency, which is particularly valuable in satellite communications, where every round trip carries a significant cost, while achieving the highest level of cryptographic assurance in both phases of the handshake.

    \subsection{LW3: KEM-based minimal exchange}

    LW3 represents the most aggressive optimization among the proposed variants, designed with the explicit goal of minimizing both communication cost and latency. To achieve this, the variant departs significantly from the standard IKEv2 design philosophy by eliminating all cryptographic negotiation: rather than dynamically agreeing on algorithms through the \texttt{SAi}/\texttt{SAr} payloads, the cipher suite is fixed and known to both peers in advance. 
    This \textit{opinionated} approach removes the overhead due to proposal exchange and selection. Although this sacrifices cryptographic agility, it is a deliberate trade-off for applications where latency and performance are critical factors. The key exchange is built entirely around static KEM public keys, pre-shared via out-of-band mechanisms prior to any protocol execution. By adopting Classic McEliece as the KEM primitive, the ciphertext remains remarkably compact across all security levels, making it one of the most bandwidth-efficient choices available. The large public key size that characterizes Classic McEliece is entirely absorbed by the out-of-band pre-sharing step, and no key material is transmitted in-band. Authentication is implicit: the ability to correctly decapsulate the ciphertext demonstrates knowledge of the corresponding static secret key, providing peer authentication without the need for a dedicated \texttt{AUTH} exchange or digital signatures. The resulting exchange, illustrated in Figure~\ref{fig:lw3_exchange}, is reduced to two messages with minimal payload, making LW3 the most bandwidth-efficient and latency-optimal configuration among the proposed variants. This, however, comes at the cost of three limitations. First, updates to static public keys must be handled entirely through out-of-band mechanisms, with no protocol-level support for key rotation. Second, the use of static key material means that Perfect Forward Secrecy (PFS) is not achieved: should a static secret key be compromised, all past sessions established using that key become retrospectively vulnerable. Third, the substantial modifications to the core exchange structure mean that backward compatibility with standard IKEv2 implementations is not preserved.

    \input{figures/lw3_exchange}

    \section{Experimental assessment}

    To evaluate the performance of the proposed variants, we set up a virtualized testbed designed to simulate different network topologies in a repeatable and fully monitored manner, enabling systematic data collection across all configurations and link characteristics. The underlying hardware platform is based on Raspbery Pi4 Model B with $8$ GB RAM. The satellite network topology considered is illustrated in Figure~\ref{fig:satellite-arch}. 

    \input{figures/testbed_arch}

    The two endpoints act as initiator and responder in the IKEv2 protocol, each connected to a modem serving as the interface to the satellite network.
    We assume that both endpoints of the communication are located on the ground and communicate via various types of satellite links.
    The satellite network, in fact, can be composed of various combinations of Low Earth Orbit (LEO), Medium Earth Orbit (MEO), and Geostationary Earth Orbit (GEO) satellites, whose link characteristics are reported in Table~\ref{tab:links}. The link parameters reflect the characteristics of real-world satellite communication systems. Propagation delays are derived from the geometric distance between the communicating nodes at the respective orbital altitudes. Jitter is inversely correlated with orbital altitude: LEO satellites, being in continuous motion, introduce greater variability in propagation delay, while GEO satellites, fixed above a given point on the Earth's surface, exhibit a nearly stable link with minimal jitter. The 5~Mbps rate assigned to ground-to-satellite links reflects a conservative estimate of the throughput available to a single user terminal over a shared satellite channel, consistent with current VSAT-class modem capabilities. Inter-satellite links are assigned 10~Gbps, in line with the throughput achievable by modern Laser Inter-Satellite Links (LISLs), which exploit free-space optical communication to achieve multi-gigabit connectivity within satellite constellations.

    \input{tables/link_parms}

    The selection of cryptographic suites was guided by the considerations reported in the previous section, with the goal of achieving security levels 1 and 3 as defined by NIST (which are roughly equivalent to $128$-bit and $192$-bit security). The corresponding choices are detailed in Table~\ref{tab:ike_variants_details}, along with a summary of the features of all the protocol variants considered.

    \input{tables/cipher_suites}

    \begin{figure}[hb]
        \centering
        \input{figures/chart/exec_time}
        \caption{Average time to complete the protocol for each variant, over 30 runs.}
        \label{fig:graph_auth}
    \end{figure}

    Figure~\ref{fig:graph_auth} reports the overall protocol execution time for all considered variants and satellite network topologies. Note that TB1 serves solely as a reference, representing the standard deployment cost against which the higher security introduced by the lightweight variants can be measured. If we consider the LEO scenario, the differences between protocols are limited: the relatively low propagation delay means that the cost of an additional round trip does not dominate the overall handshake time. The picture changes substantially in MEO and GEO scenarios, where higher propagation delays amplify the impact of each additional exchange, and the gap between variants requiring three and two messages becomes significant.

    As expected, TB2 achieves the shortest time across all scenarios; however, this comes at the cost of the limitations discussed in Section~\ref{sec:tb2}. Some of the proposed lightweight variants achieve times comparable to or smaller than TB2, while preserving full hybrid cryptography for both phases. 
    If we look at execution times, LW3 does not seem to offer any advantages over the other variants. However, if we consider the communication cost, as shown in Figure \ref{fig:comm_cost}, the picture changes significantly, with LW3 exhibiting a consistently smaller footprint than the other variants. Due to space constraints, a more comprehensive assessment that also takes memory usage and CPU time into account is left for future work.

    \input{figures/chart/communication_cost}

    \section{Conclusion}
    
    We have introduced some IKE protocol variants that appear promising for use in satellite networks. Among them, LW2 achieves hybrid key exchange and authentication in the same amount of time required by non-hybrid base configurations like TB2, while LW3 is the most efficient in communication cost. Future developments may involve optimization of cryptographic primitives for the specific case of using only ephemeral keys and the introduction of KEM-based authentication.

    \section*{Acknowledgment}

    The authors would like to thank Antonios Atlasis, Enrico Bassetti and Larissa Schrempp from the European Space Agency for their guidance and constructive feedback.
    
    \bibliographystyle{IEEEtran}
    \bibliography{bibtex/bib/rfc}

\end{document}

%% file: figures/ike_exchange.tex
    \begin{tikzpicture}[node distance=1.2cm]

        \node (entity1) [draw=ForestGreen, rectangle, thick] 
            {Initiator (\textit{i})};
        \node (entity2) [draw=Blue, rectangle, thick, 
            right=of entity1, xshift=3cm] 
            {Responder (\textit{r})};

        \draw[dashed] (entity1) -- ++(0,-5.2);
        \draw[dashed] (entity2) -- ++(0,-5.2);

        \node[font=\footnotesize\bfseries, gray] 
            at ($(entity1)!0.5!(entity2) + (0,-0.6)$) 
            {IKE\_SA\_INIT};

        \draw[-stealth, ForestGreen, thick] 
            (entity1) ++(0,-1.3) -- (entity2 |- ,-1.3) 
            node[midway, above, text=ForestGreen, font=\footnotesize] 
            {\texttt{HDR, SAi1, KEi, Ni}};

        \draw[stealth-, Blue, thick] 
            (entity1) ++(0,-2) -- (entity2 |- ,-2) 
            node[midway, above, text=Blue, font=\footnotesize] 
            {\texttt{HDR, SAr1, KEr, Nr}};

        \draw[dotted, gray] 
            ($(entity1) + (-0.5,-2.5)$) -- ($(entity2) + (0.5,-2.5)$)
            node[midway, below, font=\scriptsize, gray] 
            {encrypted from here $\longrightarrow$};

        \node[font=\footnotesize\bfseries, gray] 
            at ($(entity1)!0.5!(entity2) + (0,-3.1)$) 
            {IKE\_AUTH};

        \draw[-stealth, ForestGreen, thick] 
            (entity1) ++(0,-4) -- (entity2 |- ,-4) 
            node[midway, above, text=ForestGreen, font=\footnotesize] 
            {\texttt{HDR, SK\{IDi, AUTH, SAi2, TSi, TSr\}}};

        \draw[stealth-, Blue, thick] 
            (entity1) ++(0,-4.7) -- (entity2 |- ,-4.7) 
            node[midway, above, text=Blue, font=\footnotesize] 
            {\texttt{HDR, SK\{IDr, AUTH, SAr2, TSi, TSr\}}};

        \drawkey{0,-5.5}{orange}{\texttt{SK\_d}};
        \drawkey{6.2,-5.5}{orange}{\texttt{SK\_d}};

    \end{tikzpicture}
   

%% file: tables/ike_payloads.tex
\renewcommand{\arraystretch}{1.2}
\begin{table}[htbp]
    \centering    
    \caption{IKEv2 Payload Types}
    \label{tab:ike_payloads}
    \begin{tabular}{@{}p{1.3cm} p{6.8cm}@{}}
        \toprule
        \textbf{Name} & \textbf{Semantic} \\
        \midrule
        \texttt{HDR}      & Header of the message (not a payload). \\
        \texttt{SA}       & Security Association parameters (algorithms, DH groups).    \\
        \texttt{KE}       & Public key material for the key establishment protocol.      \\
        \texttt{N}        & Nonces selected by the initiator and responder. \\
        \texttt{ID}       & Identities to be used in authenticating the peer. \\
        \texttt{[CERT]}     & Public key certificates for authentication.                                 \\
         \texttt{[CERTREQ]}     &   Inicates the certificate authorities trusted for authentication.                        \\
        \texttt{AUTH}     & Payload that will be used to check the identity.                \\
        \texttt{TS}       & Traffic Selectors (addresses, ports).                           \\
        \texttt{SK\{..\}} & Symmetric authenticated encryption function.\\
        \bottomrule
    \end{tabular}
\end{table}

%% file: tables/ike_variants.tex
\begin{table*}[ht]
    \centering
    \caption{Summary of IKEv2 variant configurations for satellite communications}
    \label{tab:variants_summary}
    \resizebox{\textwidth}{!}{%
    \begin{tabular}{c l c l l c c l c}
    \toprule
        \textbf{Variant}                                    & 
        \makecell{\textbf{Exchanges}}                       & 
        \makecell{\textbf{Post}\\\textbf{Quantum}}          & 
        \makecell{\textbf{Hybridization}}                   & 
        \makecell{\textbf{Authentication} \textbf{Method}}  & 
        \makecell{\textbf{ROHC}}                            & 
        \makecell{\textbf{Bandwidth} \textbf{Impact}}       &
        \makecell{\textbf{PFS}}                             &
        \makecell{\textbf{IKE Compliance}}
        \\
    \midrule
        \textbf{TB1}                                            & 
            \texttt{INIT} + \texttt{INTE} + \texttt{AUTH}       & 
            \half & RFC~9370 \cite{rfc9370}                     & 
            Traditional DSS Raw Public Keys                         & 
            \xmark                                              & 
            High                                                & 
            \cmark                                              &
            \cmark

        \\[6pt]
        
        \textbf{TB2}                                            & 
            \texttt{INIT} + \texttt{AUTH}                       & 
            \full                                               & 
            \xmark                                              & 
            Symmetric PSK                                       & 
            \xmark                                              &
            Minimum                                             &
            \cmark                                              &
            \cmark                          
        \\[6pt]

        \textbf{LW1}                                            & 
            \texttt{INIT} + \texttt{INTE} + \texttt{AUTH}       & 
            \full                                               & 
            RFC 9370 \cite{rfc9370}                             & 
            PQ DSS X.509 Certificate                                  & 
            \cmark                                              & 
            Medium                                              &
            \cmark                                              &
            \cmark
        \\[6pt]

        \textbf{LW2}                                            & 
            \texttt{INIT} + \texttt{AUTH}                       & 
            \full & 
            Crockett et al. \cite{stebila-prototyping-ssh-tls}  & 
            PQ/T DSS Raw Public Keys                                      & 
            \cmark                                              &
            Low                                                 &
            \cmark                                              &
            \xmark
        \\[6pt]

        \textbf{LW3}                                            & 
            \texttt{INIT} + \texttt{AUTH}                   & 
            \full                                               & 
            \xmark                                              & 
            Implicit via out-of-band                            & 
            \cmark                                              & 
            Very low                                            &
            \xmark                                              &
            \xmark
        \\
    \bottomrule
    \\[0.05cm]
    \multicolumn{8}{l}{\small \full~Post-quantum in both INIT and AUTH \quad \half~Post-quantum in INIT only \quad \Circle~No post-quantum}

    \end{tabular}}
\end{table*}

%% file: figures/lw3_exchange.tex
\begin{figure}[htbp!]
    \centering
    \begin{tikzpicture}[node distance=0.4cm]

        \node (entity1) [draw=blue, rectangle, thick, 
            font=\small] 
            {Initiator (\textit{A})};
        \node (entity2) [draw=orange, rectangle, thick, 
            right=of entity1, xshift=0.8cm,
            font=\small] 
            {Responder (\textit{B})};

        \draw[dashed] (entity1) -- ++(0,-4.1);
        \draw[dashed] (entity2) -- ++(0,-4.1);

        \node[anchor=east, font=\scriptsize, align=right] 
            at ($(entity1) + (0,-0.8)$) 
            {$n_A \xleftarrow{\$} \{0,1\}^\lambda$};
        \node[anchor=east, font=\scriptsize, align=right] 
            at ($(entity1) + (0,-1.3)$) 
            {$k_A, c_A = \texttt{Enc}(pk_B)$};
        \node[anchor=east, font=\scriptsize, align=right] 
            at ($(entity1) + (0,-1.8)$) 
            {$x_A = \texttt{AES}(n_A, k_A)$};

        \node[anchor=west, font=\scriptsize, align=left] 
            at ($(entity2) + (0,-0.8)$) 
            {$n_B \xleftarrow{\$} \{0,1\}^\lambda$};
        \node[anchor=west, font=\scriptsize, align=left] 
            at ($(entity2) + (0,-1.3)$) 
            {$k_B, c_B = \texttt{Enc}(pk_A)$};
        \node[anchor=west, font=\scriptsize, align=left] 
            at ($(entity2) + (0,-1.8)$) 
            {$x_B = \texttt{AES}(n_B, k_B)$};

        \draw[-stealth, blue, thick] 
            ($(entity1) + (0,-2.2)$) -- ($(entity2) + (0,-2.2)$)
            node[midway, above, font=\scriptsize, text=black] 
            {$c_A,\ x_A$};

        \draw[stealth-, orange, thick] 
            ($(entity1) + (0,-2.7)$) -- ($(entity2) + (0,-2.7)$)
            node[midway, above, font=\scriptsize, text=black] 
            {$c_B,\ x_B$};

        \node[anchor=east, font=\scriptsize, align=right] 
            at ($(entity1) + (0,-3.2)$) 
            {$k_B = \texttt{Dec}(c_B, sk_A)$};
        \node[anchor=east, font=\scriptsize, align=right] 
            at ($(entity1) + (0,-3.7)$) 
            {$n_B = \texttt{AES}^{-1}(x_B, k_B)$};

        \node[anchor=west, font=\scriptsize, align=left] 
            at ($(entity2) + (0,-3.2)$) 
            {$k_A = \texttt{Dec}(c_A, sk_B)$};
        \node[anchor=west, font=\scriptsize, align=left] 
            at ($(entity2) + (0,-3.7)$) 
            {$n_A = \texttt{AES}^{-1}(x_A, k_A)$};

        \node[anchor=east, font=\scriptsize, align=right] 
            at ($(entity1) + (4,-4.5)$) 
            {$\texttt{SKEYSEED} = \texttt{PRF}(n_A \| n_B,\ k_A \| k_B)$};

    \end{tikzpicture}
    \caption{LW3 exchange flow. Both peers encapsulate a fresh 
    nonce under the other's static KEM public key, pre-shared 
    via out-of-band mechanisms. Authentication is implicit: 
    successful decapsulation proves knowledge of the 
    corresponding static secret key.}
    \label{fig:lw3_exchange}
\end{figure}

%% file: figures/testbed_arch.tex
\begin{figure}[hb]
\centering
\begin{tikzpicture}[
    node distance=0.6cm and 0.8cm,
    box/.style={
        rectangle, rounded corners=3pt,
        draw=black, thick,
        minimum width=1.6cm, minimum height=0.7cm,
        text centered, font=\small
    },
    satnode/.style={
        font=\small,
        text centered
    },
    netbox/.style={
        rectangle, rounded corners=6pt,
        draw=black!60, thick, dashed,
        fill=blue!5,
        inner sep=8pt
    },
    darr/.style={{Stealth[length=4pt]}-{Stealth[length=4pt]}, thick},
]

\newcommand{\saticon}[1]{%
    \begin{tikzpicture}[scale=0.32, baseline=-0.1cm]
        \draw[fill=gray!50, draw=black!70, thick, rounded corners=1pt]
            (-0.35,-0.28) rectangle (0.35,0.28);
        \draw[gray!80, thin] (-0.35,0) -- (0.35,0);
        \draw[gray!80, thin] (0,-0.28) -- (0,0.28);

        \draw[fill=blue!50, draw=black!70] (-1.15,-0.20) rectangle (-0.35,0.20);
        \foreach \x in {-0.97,-0.78,-0.59} {
            \draw[black!40, very thin] (\x,-0.20) -- (\x,0.20);
        }
        \draw[black!40, very thin] (-1.15,0.03) -- (-0.35,0.03);
        \draw[black!40, very thin] (-1.15,-0.08) -- (-0.35,-0.08);

        \draw[fill=blue!50, draw=black!70] (0.35,-0.20) rectangle (1.15,0.20);
        \foreach \x in {0.53,0.72,0.91} {
            \draw[black!40, very thin] (\x,-0.20) -- (\x,0.20);
        }
        \draw[black!40, very thin] (0.35,0.03) -- (1.15,0.03);
        \draw[black!40, very thin] (0.35,-0.08) -- (1.15,-0.08);

        \draw[fill=white!90!gray, draw=black!70]
            plot[domain=-0.22:0.22, samples=20]
            (\x, {0.28 + 0.5*\x*\x + 0.38});
        \draw[black!70] (0, 0.28) -- (0, 0.66);   
        \draw[fill=black!60, draw=black!70] (0, 0.66) circle (0.025); 

        \draw[black!70, thin] (0.18,-0.28) -- (0.25,-0.50);
        \draw[fill=black!50] (0.25,-0.50) circle (0.02);
    \end{tikzpicture}%
    \;\,#1%
}

\node[satnode] (geo) {\saticon{GEO}};
\node[satnode, below left=0.55cm and 0.5cm of geo]  (leo) {\saticon{LEO}};
\node[satnode, below right=0.25cm and 0.5cm of geo] (meo) {\saticon{MEO}};

\begin{scope}[on background layer]
    \node[netbox, fit=(geo)(leo)(meo),
          label={[font=\small\itshape]above:Satellite Network}] (satnet) {};
\end{scope}

\node[box, fill=orange!15, below left=1cm and 1.6cm of satnet.south] (modemA)   {Modem A};
\node[box, fill=green!15,  below=0.6cm of modemA]                      (initiator) {Initiator};

\node[box, fill=orange!15, below right=1cm and 1.6cm of satnet.south] (modemB)   {Modem B};
\node[box, fill=green!15,  below=0.6cm of modemB]                       (responder) {Responder};

\draw[dashed, thick, gray!50, dash pattern=on 6pt off 4pt]
    (current bounding box.west |- 0,-2.3) -- 
    (current bounding box.east |- 0,-2.3);
\node[font=\footnotesize\itshape, text=gray!60, anchor=south]
    at (current bounding box.center |- 0,-2.3) {Space segment};
\node[font=\footnotesize\itshape, text=gray!60, anchor=north]
    at (current bounding box.center |- 0,-2.3) {Ground segment};

\draw[darr] (initiator) -- (modemA);
\draw[darr] (modemB)    -- (responder);
\draw[darr] (modemA.north) -- (satnet.south -| modemA);
\draw[darr] (modemB.north) -- (satnet.south -| modemB);

\end{tikzpicture}
\caption{Communication architecture through a satellite network.}
\label{fig:satellite-arch}
\end{figure}

%% file: tables/link_parms.tex
\begin{table}[t!]
\caption{Characteristics of considered communication links.}
\centering
\renewcommand{\arraystretch}{1.2} 
\begin{tabular}{ l c l c r r }
\toprule
\textbf{Link} & & & \textbf{Rate} & \textbf{Delay} & \textbf{Jitter} \\
\midrule
Host   & $\leftrightarrow$ & Modem & $\infty$   & \num{22.500} ms  & \num{0}                \\
Modem  & $\leftrightarrow$  & LEO   & 5~Mbps    & \num{6.862} ms   & \num{1.178} ms        \\
Modem  & $\leftrightarrow$  & MEO   & 5~Mbps    & \num{78.915} ms  & \num{0.14} ms         \\
Modem  & $\leftrightarrow$  & GEO   & 5~Mbps    & \num{127.247} ms & \num{0.04} $\mu$s         \\
LEO    & $\leftrightarrow$  & LEO   & 10~Gbps   & \num{15.898} ms  & \num{2.73} ms         \\
LEO    & $\leftrightarrow$  & MEO   & 10~Gbps   & \num{41.167} ms  & \num{0.073} ms        \\
GEO    & $\leftrightarrow$  & MEO   & 10~Gbps   & \num{92.500} ms  & \num{0.029} ms        \\
\bottomrule
\end{tabular}
\label{tab:links}
\end{table}

%% file: tables/cipher_suites.tex
\begin{table*}[!ht]
\caption{Cryptographic suites for the variants.}
\label{tab:ike_variants_details}
\centering
\renewcommand{\arraystretch}{1.15} 
\begin{tabular}{c l c c c l l l}
\toprule
    \textbf{Security Level} & 
    \textbf{Variant}        & 
    \textbf{Sym. cipher}    & 
    \textbf{Hash}           & 
    \textbf{Trad. KEX}      & 
    \textbf{PQC KEM}        &
    \textbf{Trad. Auth}     &
    \textbf{PQC Auth} \\
\midrule

\multirow{5}{*}{I} 
& TB1                                      & 
    \multirow{5}{*}{\texttt{AES-128-GCM}}   &  
    \multirow{5}{*}{\texttt{SHA-256}}       & 
    \texttt{x25519}                         & 
    \texttt{ML-KEM-512}                     &    
    \texttt{ECDSA}                          &
    \xmark
\\

& TB2                                      &                               
                                            &                           
                                            & 
    \xmark                                  & 
    \texttt{ML-KEM-512}                     & 
    \texttt{Symmetric PSK}                  &
    \xmark
\\
   
& LW1                                       &
                                            &
                                            &
    \texttt{x25519}                         & 
    \texttt{ML-KEM-512}                     & 
    \xmark                                  &
    \texttt{ML-DSA-44}
\\
& LW2                                       &                               
                                            &
                                            & 
    \texttt{x25519}                         & 
    \texttt{ML-KEM-512}                     & 
    \texttt{ECDSA}                          &  
    \texttt{ML-DSA-44} 
\\
& LW3                                       &                               
                                            &                           
                                            & 
    \xmark                                  & 
    \texttt{McEliece-348864}                & 
    \xmark                                  &
    -
\\[0.1cm]
\hline
\addlinespace[0.1cm]
\multirow{5}{*}{III} 
& TB1                                       & 
    \multirow{5}{*}{\texttt{AES-192-GCM}}   &  
    \multirow{5}{*}{\texttt{SHA-384}}       & 
    \texttt{ECP384}                         & 
    \texttt{ML-KEM-768}                     &    
    \texttt{ECDSA}                          &
    \xmark
\\
& TB2                                       &                               
                                            &                           
                                            & 
    \xmark                                  & 
    \texttt{ML-KEM-768}                     & 
    \texttt{Symmetric PSK}                  &
    \xmark
\\
& LW1                                       &                               
                                            &                           
                                            & 
    \texttt{ECP384}                        & 
    \texttt{ML-KEM-768}                     & 
    \xmark                                  &
    \texttt{ML-DSA-65}
\\
& LW2                                       & 
                                            &  
                                            & 
    \texttt{ECP384}                         & 
    \texttt{ML-KEM-768}                     &    
    \texttt{ECDSA}                          & 
    \texttt{ML-DSA-65} 
\\
& LW3                                       &                               
                                            &                           
                                            & 
    \xmark                                  & 
    \texttt{McEliece-460896}                & 
    \xmark                                  &
    -
\\

\bottomrule
\end{tabular}
\end{table*}

%% file: figures/chart/exec_time.tex
\begin{tikzpicture}
\begin{axis}[
    ybar,
    bar width=4pt,
    width=\linewidth,
    height=6cm,
    ylabel={Overall execution time (s)},
    symbolic x coords={LEO, MEO, GEO},
    xtick=data,
    x tick label style={font=\footnotesize},
    enlarge x limits=0.3,
    ymin=0,
    ymax=2,
    ytick={0,0.2,0.4,0.6,0.8,1,1.2,1.4,1.6,1.8},
    y tick label style={font=\footnotesize},    
    y label style={font=\footnotesize},    
    ymajorgrids=true,
    grid style={dashed,gray!50},
    major grid style={draw=gray!50},
    every axis plot/.append style={draw=none},
    every node near coord/.append style={font=\tiny},
    legend style={
        at={(0.5,1.02)},
        anchor=south,
        legend columns=5,
        column sep=0.5ex,
        draw=none,
        font=\tiny
    },
]
\addplot[fill=tbone] plot coordinates {(LEO,0.2908) (MEO,1.1515) (GEO,1.7321)};
\addplot[fill=tbone!65!black] plot coordinates {(LEO,0.2949) (MEO,1.1777) (GEO,1.7557)};
\addplot[fill=tbtwo] plot coordinates {(LEO,0.18397) (MEO,0.7656) (GEO,1.1444)};
\addplot[fill=tbtwo!65!black] plot coordinates {(LEO,0.2817) (MEO,1.1579) (GEO,1.7319)};
\addplot[fill=lwone] plot coordinates {(LEO,0.304023) (MEO,1.169918) (GEO,1.738271)};
\addplot[fill=lwone!70!black] plot coordinates {(LEO,0.329887) (MEO,1.209255) (GEO,1.782741)};
\addplot[fill=lwtwo] plot coordinates {(LEO,0.1895) (MEO,0.7610) (GEO,1.1449)};
\addplot[fill=lwtwo!65!black] plot coordinates {(LEO,0.1948) (MEO,0.7610) (GEO,1.1449)};
\addplot[fill=lwthree] plot coordinates {(LEO,0.2850) (MEO,0.8682) (GEO,1.2541)};
\addplot[fill=lwthree!65!black] plot coordinates {(LEO,0.2850) (MEO,0.8682) (GEO,1.2541)};
\legend{
TB1-I, TB1-III,
TB2-I, TB2-III,
LW1-I, LW1-III,
LW2-I, LW2-III,
LW3-I, LW3-III
}
\end{axis}
\end{tikzpicture}

%% file: figures/chart/communication_cost.tex
\begin{figure}[ht]
\centering
\begin{tikzpicture}
\begin{axis}[
    nodes near coords,
    nodes near coords align={vertical},
     every node near coord/.append style={
        font=\tiny, 
    },
    ybar,
    bar width=0.3cm,
    width=0.9\columnwidth,
    height=6cm,
    legend style={
        at={(0.5,1.05)},
        anchor=south,
        legend columns=2,
        draw=none,
        font=\footnotesize
    },
    symbolic x coords={TB1, TB2, LW1, LW2, LW3},
    xtick=data,
    x tick label style={font=\small},
    y tick label style={font=\footnotesize},
    ylabel style={font=\footnotesize},
    ymin=0,
    ymax=12000,
    ytick={0, 2500, 5000, 7500, 10000, 12500, 15000, 17500, 20000, 22500, 25000},
    scaled y ticks=false,
    ylabel={Communication cost (bytes)},
    ylabel style={font=\footnotesize, yshift=2pt},
    enlarge x limits=0.2,
    grid=major,
    grid style={dashed, gray!30},
]

\addplot[fill=blue!40, draw=blue!70!black , bar shift=-9pt] 
    coordinates {
    (TB1,2826) 
    (TB2,2509) 
    (LW1,7975) 
    (LW2,7939) 
    (LW3,725)};

\addplot[fill=orange!40, draw=orange!70!black] 
    coordinates {
    (TB1,3722) 
    (TB2,3245) 
    (LW1,10643) 
    (LW2,10715) 
    (LW3,845)};

\legend{128-bit security, 192-bit security}

\end{axis}
\end{tikzpicture}
\caption{Total communication cost for the considered protocol variants.}
\label{fig:comm_cost}
\end{figure}
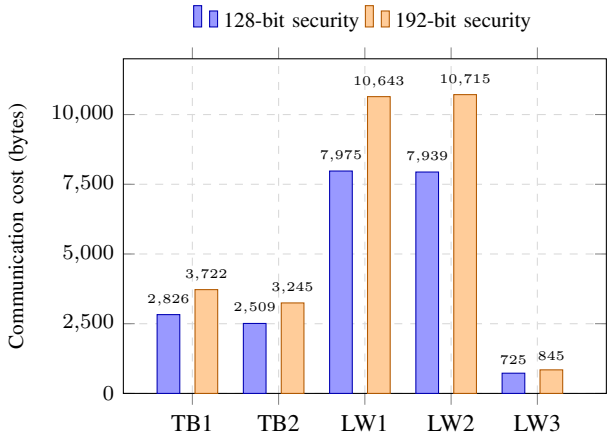